\documentclass{article} 
\usepackage{iclr2027_conference,times}

\usepackage{amsmath,amsfonts,bm}

\def\eqref#1{equation~\ref{#1}}

\def\1{\bm{1}}

\DeclareMathAlphabet{\mathsfit}{\encodingdefault}{\sfdefault}{m}{sl}
\SetMathAlphabet{\mathsfit}{bold}{\encodingdefault}{\sfdefault}{bx}{n}

\iclrfinalcopy

\usepackage{hyperref}
\usepackage{url}

\usepackage{booktabs}       
\usepackage{amsfonts}       
\usepackage{nicefrac}       
\usepackage{microtype}      
\usepackage{xcolor}         

\usepackage{enumitem}
\usepackage{amsmath}
\usepackage{float}
\usepackage{graphicx}
\usepackage{svg}
\usepackage{multirow}
\usepackage[ruled,vlined,linesnumbered]{algorithm2e}

\title{FedJigsaw: Multi-Agent Collaborative Model Reassembly for Decentralized Heterogeneous Federated Learning}

\author{%
  Jifeng Chen, Haibo Zhang, Yawen Chen \\
  School of System \& Computing\\
  University of New South Wales\\
  Canberra ACT 2600 Australia \\
  \texttt{\{jifeng.chen, haibo.zhang, wendy.chen1 \}@unsw.edu.au} \\
}

\begin{document}

\maketitle

\begin{abstract}
Model Heterogeneous Federated Learning (MHFL) addresses client-level resource heterogeneity by allowing each participant to train a personalized model architecture under a shared training objective. A prevalent paradigm, Partial Training (PT), achieves this by allowing each client to train a subnetwork of the global model. However, existing PT methods typically rely on predefined architectural templates or over-parameterized supernets, limiting fine-grained personalization and imposing substantial computational and memory overhead. We propose \texttt{FedJigsaw}, a novel framework that reshapes model personalization as a dynamic and decentralized model assembly problem. Instead of selecting subnetworks from a predefined supernetwork, each client constructs its model by assembling reusable modules learned from neighboring clients. 
At the client level, we introduce \texttt{AttenAssemble} to enable each participant to adaptively construct a tailored model based on local observations. To support efficient knowledge sharing under communication and privacy constraints, we design \texttt{SymbioArchitect}, a mechanism that allows clients to exchange granular model modules with their topological neighbors. To mitigate training instability introduced by decentralized module exchange,  we design \texttt{CoRe-Tune}, an attention-enhanced centralized training with a decentralized execution strategy, which guides local policies to foster implicit collaboration and stabilize training dynamics, without compromising data privacy. Extensive evaluations demonstrate that \texttt{FedJigsaw} outperforms state-of-the-art MHFL baselines by up to 13.8\% in relative accuracy while significantly shrinking cross-client performance variance, but also slashes decision-making latency and peak memory footprint compared to existing policy-driven methods. Ablation studies show that dynamic modular composition is more effective and efficient than supernet-based partial training under heterogeneous settings.
\end{abstract}

\section{Introduction}

Federated Learning (FL) has emerged for collaborative model training while preserving data privacy. However, when deploying FL into real-life scenarios, it inevitably encounters data, computational, and communicational heterogeneity \citep{FLSurvey}. Early efforts \citep{FedPAC, tastan2024redefining, li2021model} primarily focused on maintaining personalized parameters or customized aggregating weights. While early efforts are effective for data heterogeneity, computational heterogeneity is overlooked. Essentially, no weight tuning can magically enable a resource-constrained microcontroller to infer a ResNet-50 model in an FL system.





To bridge this gap, Model Heterogeneous Federated Learning (MHFL) has been proposed. Despite recent advances, current mainstream MHFL methodologies remain fundamentally bottlenecked by the following three main aspects:
\begin{itemize}[leftmargin=*]
    \item \textbf{Heuristic-driven Inflexibility:} Existing methods\citep{alam2022fedrolex, horvath2021fjord, chen2024flexfl} typically rely on static sub-networks extracted to a few discrete predefined capability levels. Consequently, they fail to customize architectures for individual client capacities and preclude true architectural diversity by strictly requiring a unified backbone.

    \item \textbf{Resource-prohibitive overhead:} Another MHFL branch leverages Knowledge Distillation (KD)\citep{wang2023knowledge,FedMRL}. However, these methods require clients to maintain an auxiliary homogeneous or teacher model alongside the local model, which inherently doubles the local memory and computational overhead, making them impractical for resource-constrained edge devices.

    \item \textbf{Privacy-intrusive centralization:} Several MHFL approaches \citep{lin2020ensemble,li2019fedmd} rely on public datasets as a shared reference medium for feature alignment and knowledge transfer. Although this strategy facilitates global consensus, it inevitably introduces massive computational overhead. More critically, sharing inference results on public data reintroduces severe privacy vulnerabilities, fundamentally contradicting the premise of FL.
    
\end{itemize}


        
        

\paragraph{Motivation \& Core Challenges}

Motivated by these limitations, an ideal MHFL paradigm should empower each client to dynamically and autonomously assemble a customized architecture tailored to its unique context. Furthermore, the architectural decision-making processes must be fully decentralized. By relying on local agents rather than centralized public datasets or extensive server-side coordination, such a paradigm inherently mitigates privacy risks and minimizes deployment costs. Realizing this motivation, however, poses three core challenges:

\begin{itemize}[leftmargin=*]
    \item \textbf{C1: Efficient Context-Aware Architecture Generation:} Traditional Neural Architecture Search (NAS) is computationally expensive and requires iterative evaluations, making it challenging for real-time execution or on resource-constrained edge devices. Achieving rapid, context-aware model generation, where a client can direct the "assembly" of a customized model with low inference latency from its local data features, remains a critical hurdle.
    
    \item \textbf{C2: Structural Knowledge Sharing Beyond Parameters:} Conventional FL systems facilitate knowledge sharing strictly through the aggregation of model parameters. In an MHFL environment, client model backbones may also diverge; this parameter-level communication completely breaks down. In a decentralized MHFL environment, how to break this physical isolation and establish a collaborative scheme for sharing structural blocks becomes a fundamental problem.
    
    \item \textbf{C3: Stable and Efficient Optimization:} To realize dynamic assembly, each client is inherently modeled as an autonomous learning agent. However, in a multi-agent environment, as individual agents continuously update their architectural policies, environmental feedback becomes severely non-stationary from the perspective of any single agent, forming a non-cooperative game. Thus, the guidance of training for this system is necessary and challenging.
    
\end{itemize}





\paragraph{Contributions}
To address these challenges, we propose \textbf{FedJigsaw}, a novel framework for Multi-Agent collaborative reassembly in MHFL. By formulating the problem as a Multi-Agent Reinforcement Learning (MARL) paradigm, \textbf{each client possesses an agent}, which dynamically reassembles neural networks block by block locally, similar to solving a jigsaw puzzle.
\begin{itemize}[leftmargin=*]
    \item \textbf{Solution to C1:} To address C1, we propose the \textbf{AttenAssemble} paradigm. Unlike costly global supernet searches in NAS, we formulate model assembly as a sequential decision-making process. Powered by an attention-based pointer network, this paradigm rapidly identifies the correlations between candidate actions and the current state, and dynamically selects the optimal modules.
    
    \item \textbf{Solution to C2:} To address C2, we design the \textbf{SymbioArchitect} scheme to enable knowledge transfer across heterogenenous models. Instead of exchanging full models, clients collaboratively fetch specific neural modules from their neighbors guided by lightweight metadata. Consequently, this not only elevates knowledge to a structural level but also drastically reduces communication overhead by transmitting only the necessary components.

    \item \textbf{Solution to C3:} To address C3, we introduce the \textbf{CoRe-Tune} strategy to mitigate non-stationarity caused by the MARL non-cooperative game. Utilizing an attention-enhanced Centralized Training Decentralized Execution (CTDE) framework, a global \texttt{ServerCritic} leverages its global vision to explicitly guide local policy training. Once transitioned to decentralized execution, the server is entirely decoupled. This guarantees that clients' decision-making remains physically independent, yet achieves profound implicit collaboration.
\end{itemize}












\section{Literature Review}

\paragraph{Knowledge Distillation and Mutual Learning in MHFL}
To address the model-heterogeneity challenge in MHFL, a significant line of research leverages KD and Mutual Learning to transfer knowledge via a global sharing model and a local heterogeneous model, then uses parameters to propagate knowledge. Methods such as FedDF \citep{lin2020ensemble} and FedMD \citep{li2019fedmd} rely on a public dataset as a shared reference and exchange logits to reach global consensus. Alternatively, approaches such as pFedKnow \citep{wang2023knowledge} and FedMRL \citep{FedMRL} utilize pre-trained large models or enforce representational alignment between local heterogeneous models and a shared homogeneous model. However, these techniques suffer from two main limitations. First, KD-based approaches inherently require clients to maintain an extra auxiliary or teacher model, imposing prohibitive computational and memory overheads on resource-constrained edge devices\citep{FLSurvey}. Secondly, relying on public datasets for feature alignment introduces both massive inference complexity and severe privacy vulnerabilities. Consequently, these limitations render such a paradigm fundamentally impractical for resource-constrained environments.

\paragraph{Partial Training in MHFL}
Another branch achieves heterogeneity through partial training or sub-network extraction \citep{wen2022federated}. Frameworks such as HeteroFL \citep{diao2020heterofl}, FjORD \citep{horvath2021fjord}, FlexFL \citep{chen2024flexfl} extract static sub-networks based on discrete, predefined capability levels. Similarly, FedRolex \citep{alam2022fedrolex} introduces a rolling submodel extraction scheme to evenly train the global model. Based on that, FedShapleX \citep{FedShapleX} uses Shapley value to achieve a fair and context-aware submodel extraction. Although these methods have effectively reduced the computation overhead, they strictly require all clients to share a unified global supernet, precluding true architectural diversity. Furthermore, these coarse-grained extractions fail to customize architectures to individual dynamic capacities. Crucially, blind truncation of rolling layers disrupts strict mathematical alignments (e.g., residual connections), leading to semantic misalignment and performance degradation.

\paragraph{Federated Neural Architecture Search}
To achieve true structural personalization, recent efforts have integrated Neural Architecture Search (NAS) into FL. FedNAS \citep{FedNAS} jointly optimizes model weights and architecture parameters within a massive supernet, but suffers from severe communication bottlenecks. pFedHR \citep{pfedhr} attempts heterogeneous model reassembly \citep{yang2022deep}, yet introduces unacceptable computational complexity during candidate model matching and heavily relies on the public dataset. More recently, RL-based FedNAS approaches such as PerFedRLNAS\citep{yao2024perfedrlnas} and Peaches \citep{yan2024peaches} employ decision-making networks to determine whether a component is a shard or personalized. While promising, the current methods still rely on a public dataset for candidate matching and introduce a high computational cost. Thus, a promising direction would be to allow each client to explore the suitable model on their own, further improving privacy and efficiency.

\section{Methodology}
To address the inherent constraints of MHFL, we propose \textbf{FedJigsaw}, which comprises three main components: the model-assembly algorithm \textbf{AttenAssemble}, the decentralized multi-agent sharing architecture \textbf{SymbioArchitect}, and the global tuning strategy \textbf{CoRe-Tune}.

In the proposed \texttt{FedJigsaw} MHFL system, during communication round $k$, the participating clients' collective model space is represented as $\mathcal{M}_k = \left\{ \left( \theta_{k, l}^{(i)}, m_{k,l}^{(i)} \right) \mid i \in \mathcal{N}, \, l \in \{1, \dots, L_i\} \right\}$, where $\mathcal{N}$ represents the set of all clients, and $L_i$ denotes the total number of structural modules within the heterogeneous architecture of client $i$. We define a module as a cohesive, human-designed functional block that encapsulates multiple fundamental layers (e.g., ResNet residual block, Transformer Encoder block). Specifically, $\theta_{k, l}^{(i)}$ parameterizes the trainable weights of the $l$-th module on client $i$, while $m_{k,l}^{(i)} = \langle c_{id}, l_{id}, type_{module}, \mathbf{d}_{in}, \mathbf{d}_{out} \rangle$ encapsulates its associated lightweight metadata. This 5-tuple describes module identity, detailing its source client id ($c_{id}$), original layer position ($l_{id}$), operation type ($type_{module}$), and structural dimensionalities ($\mathbf{d}_{in}$, $\mathbf{d}_{out}$). In the subsequent subsections, we formally detail the execution logic of three core components within our system.

\subsection{AttenAssemble: Attention-based Model Reassembly}
\label{sec:atten}
Unlike traditional NAS methodologies that rely on computationally costly supernet extractions\citep{pham2018efficient}, our proposed \texttt{AttenAssemble} paradigm formulates dynamic model generation as a bottom-up sequential decision process. Like playing Jigsaw, at each step, the policy only needs to consider which layer should be next, and the action space in each step will be narrowed down. 

\subsubsection{Model Assembling Problem Formulation}
\label{sec:MDP}
Specifically, the agent on client $i$ firstly observes a candidate module zoo $\mathcal{Z}_k^{(i)} \subseteq \mathcal{M}_k$, which contains the modules that the agent can observe and select from. By autoregressively selecting and appending structural modules from this pool, it independently generates its customized model architecture $\Theta_k^{(i)}$ for the current communication round. We strictly map this layer-wise construction procedure into an MDP, which is formulated as follows:

\begin{itemize}[leftmargin=*]
    \item \textbf{State:} As we formulate the model assembling as a sequential decision problem, we define the system state $s_{k,t}^{(i)} = \left( E_{data}^{(i)}, \{m_{k, l_\tau}^{(\cdot)}\}_{\tau=1}^t \right), s_{k,t}^{(i)} \in \mathcal{S}$ as the intermediate topological architecture constructed up to step $t$ by client $i$ during communication round $k$, which encapsulates the unified local data feature $E_{data}^{(i)}$ and the metadata from all previously assembled modules.
    
    \item \textbf{Action:} At any step $t$, the valid action space $\mathcal{A}_{k,t}^{(i)}$ is a dynamically filtered subset of the local candidate pool $\mathcal{Z}_k^{(i)}$. An action $a_{k,t}^{(i)}=append(m_{k,l}^{(\cdot)})$ is strictly valid only if its input dimension geometrically aligns with the output dimension of the current state $s_{k,t}^{(i)}$. Furthermore, the action space incorporates a distinct \textit{STOP} action to terminate the assembly process. This early-stopping mechanism allows the agents to flexibly find the personalized architecture under current constraints, rather than exhausting all available computational resources.
    
    \item \textbf{Transition Probability:} The model structural transition in our system is strictly deterministic.
    
    \item \textbf{Reward:} Since intermediate network assemblies cannot be meaningfully evaluated, the decision-making environment operates under a strictly sparse reward setting, yielding zero rewards during the assembly phase. A terminal reward $R_{k,T}^{(i)}$ is only acquired after the entire model is constructed and evaluated. Consequently, to resolve the credit assignment problem for intermediate actions, the Monte-Carlo return is utilized, formulated as $r_{k,t}^{(i)} = \gamma^{T-t} R_{k,T}^{(i)}$.
    
\end{itemize}

\subsubsection{Attention-based Pointer Network Actor}

While Deep Reinforcement Learning (DRL) serves as a natural paradigm for generalizing across dynamic environments, standard DRL actors typically rely on fixed-dimensional action spaces. This severely limits their applicability in our formulated MDP, where the available architectural modules (i.e., action space) dynamically fluctuate based on the topological neighbors. 

To bridge this gap, inspired by the original pointer network\citep{vinyals2015pointer}, \textbf{AttenAssemble} introduces a tailored Attention-based Pointer Network Actor. By leveraging a Query-Key attention matching process, this actor effortlessly handles variable-length action spaces and captures the complex correlations between the current architecture and the candidate modules. Remarkably, despite observing only metadata, the actor implicitly learns the learned semantic compatibility through continuous environmental feedback. This attention mechanism precludes semantic mismatches, thereby circumventing the severe gradient conflicts that may happen in parameter aggregation. The pseudocode is provided in Appendix \ref{app:actor}.

\begin{figure}[htbp]
    \centering
    \includegraphics[width=0.8\linewidth]{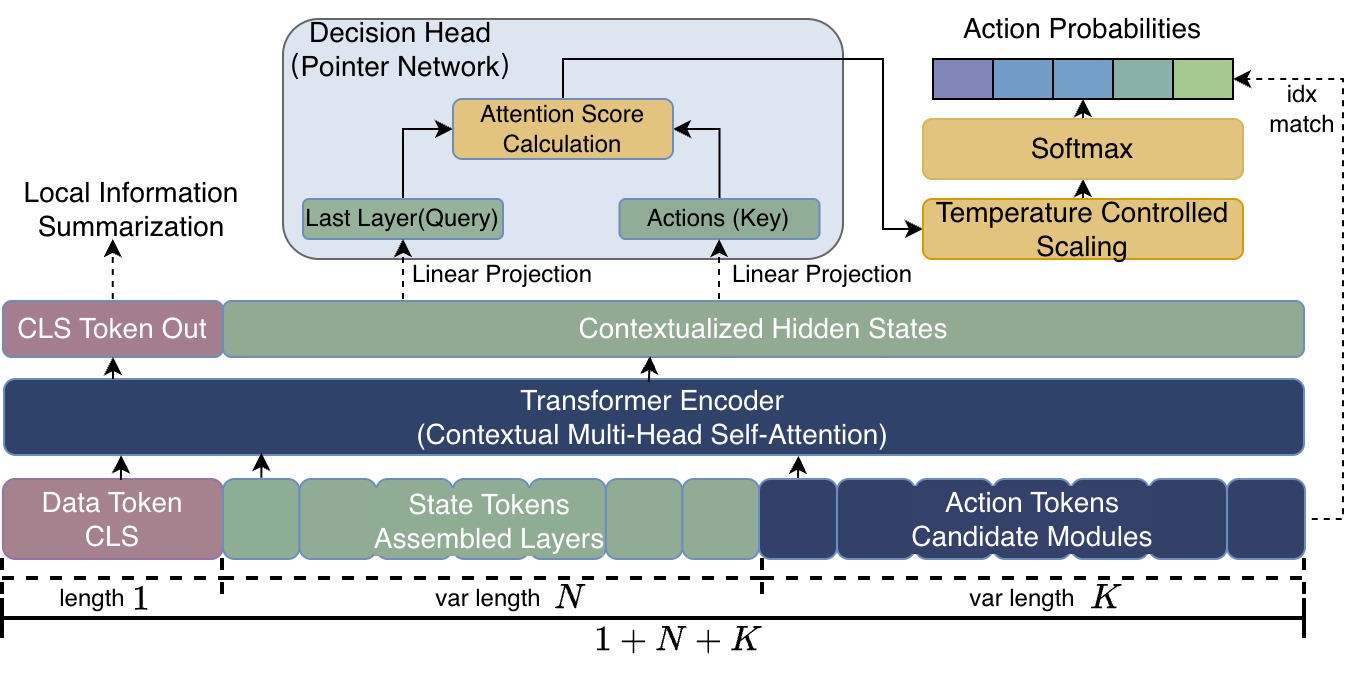}
    \vspace{-15pt}
    \caption{The architecture of the \texttt{AttenAssemble} policy network.}
    \vspace{-10pt}
    \label{fig:actor}
\end{figure}

\paragraph{Policy Network} As depicted in Fig.~\ref{fig:actor}, we formulate the model assembling process as a sequence modeling task, drawing inspiration from natural language processing. Prior to sequence construction, the raw local data representations and structural metadata are projected into a unified embedding space via lightweight encoders. Subsequently, \texttt{AttenAssemble} constructs a unified input sequence, which is concatenated by the local data feature token $E_{Data}^{(i)}$ (serving as the \texttt{CLS} token), the sequentially assembled state tokens $E_{State}$ (length $N$), and the valid candidate action tokens $E_{Action}$ (length $K$).

The sequence will then be contextualized by a multi-head Transformer encoder. To seamlessly handle the variable action space, a \texttt{Pointer Network} head is employed to compute the action distribution via dot-product attention score, which directly formulates the policy $\pi(a \mid s)$ in our reinforcement learning framework. Specifically, it utilizes the  hidden representation of the most recent assembled layer as the query $q$, and the $K$ candidates as the key matrix $K_{Act}$:
\begin{equation}
    \pi_{\theta}(a \mid s_{k,t}^{(i)}) = \text{Softmax}\left( \frac{q W_Q (K_{act} W_K)^\top}{\tau \sqrt{d_{model}}} \right)
    \label{Policy}
\end{equation}

In this way, not only the relationship between the last layer and the actions has been considered, but also the information of the previous layers has been encoded into the query $q$ through the Transformer encoder. To facilitate exploration during policy training, a temperature parameter $\tau$ is introduced to scale the raw logits generated by the Pointer Network. Applying this scaling prior to the Softmax activation dynamically modulates the entropy of the action distribution, thereby preventing premature greedy selection leading to a local optimum. Subsequently, the Actor network is optimized via the \textbf{REINFORCE} algorithm, updating the parameters in the direction of the module selection log-probabilities scaled by the discounted returns.


\begin{figure}[htbp]
    \centering
    \includegraphics[width=0.9\linewidth]{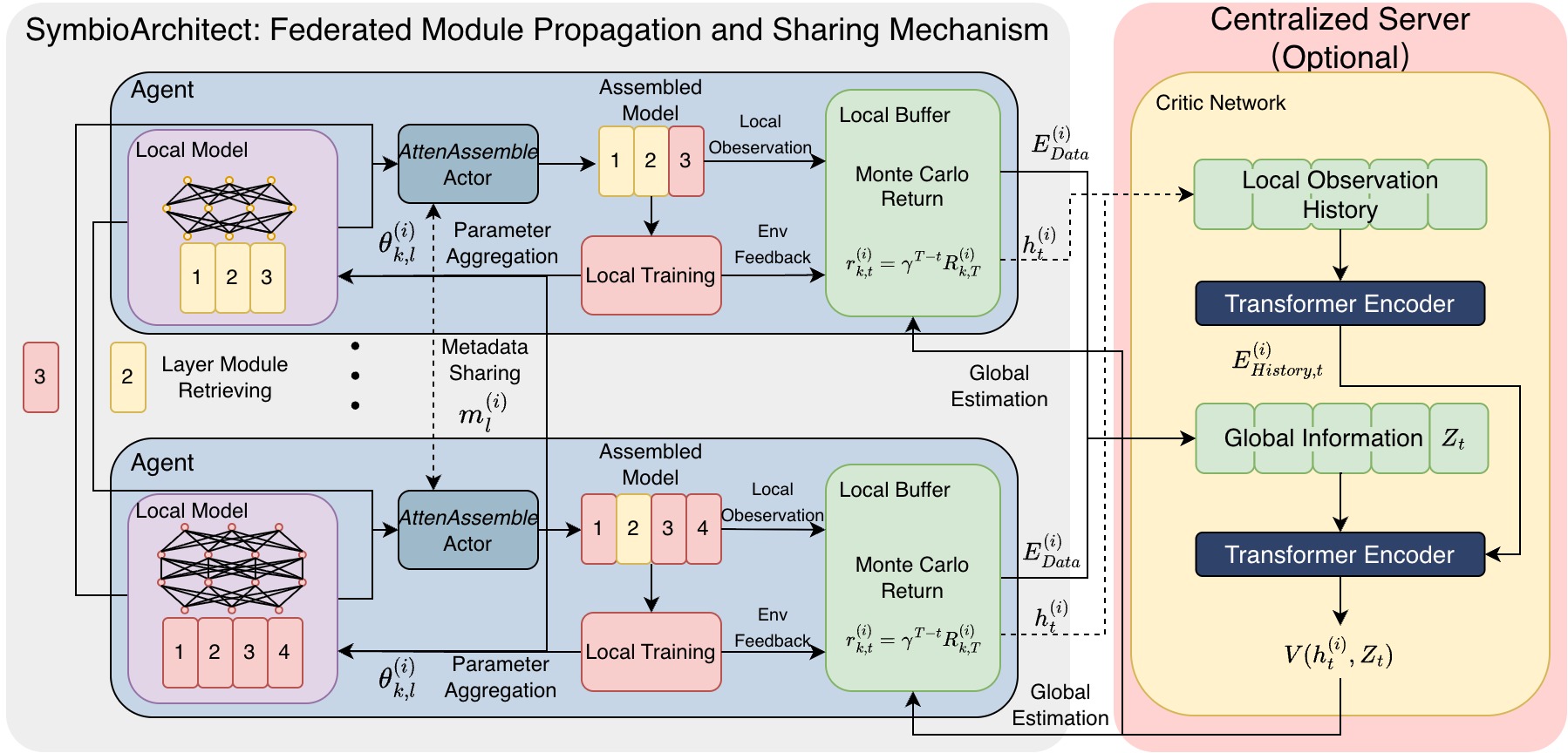}
    \vspace{-5pt}
    \caption{The architecture of the \texttt{FedJigsaw} MHFL system. The left side illustrates the \texttt{SymbioArchitect}, while the right side shows the Centralized Server used in the CTDE scheme.}
    \vspace{-10pt}
    \label{fig:FedJigsaw}
\end{figure}

\subsection{SymbioArchitect: Federated Module Propagation and Sharing Mechanism}
\label{sec:symbio}

Building upon the model assembly mechanism introduced in Section~\ref{sec:atten}, this section broadens the scope to system-wide collaboration. Specifically, we will discuss how clients dynamically exchange knowledge, including both parameters and structural models, across a decentralized MHFL network.

\subsubsection{Sharing Scheme}
\paragraph{Structural Sharing} 

Building upon this layer-wise model assembling scheme, a critical challenge lies in managing the dynamic action space. Recent approaches like pFedHR\citep{pfedhr} impose rigid heuristic constraints to ensure a lower bound of module utilization. However, under such paradigms, unselected modules in early rounds are often prematurely abandoned, leading to architectural collapse. Furthermore, their reliance on "layer stitching" to resolve dimensional mismatches frequently introduces a prohibitive parameter overhead.

To stabilize the candidate pool and prevent module deprecation, as illustrated in Fig.~\ref{fig:FedJigsaw}, each client maintains its assigned task-specific architecture as a local base model, thereby guaranteeing a structurally bounded search space. Notably, unlike existing approaches, this local base model functions exclusively as a passive knowledge repository and \textbf{doesn't participate in active computation}. Because it bypasses parameter updates during local training, it doesn't introduce additional computational overhead. Under this paradigm, the global collective model space is formally defined as $\mathcal{M}_k=\{(\theta_{k,l}^{(j)},m_{l}^{(j)})\mid j \in \mathcal{N}, l \in \{1, \dots, L_i\}\}$ as the metadata won't change.

Instead of synchronizing entire architectures\citep{pfedhr}, which results in high communicational overhead, clients merely broadcast their \textbf{structural metadata} $m_{l}^{(j)}$ within their neighborhood defined as $\mathcal{N}_i \subseteq \mathcal{N}$. The \texttt{AttenAssemble} agent screens this metadata to filter compatible candidate modules. Upon selection, the physical parameters of the target layer $\theta_{k,l}^{(j)}$  are retrieved for local assembly, and this routing decision is immediately logged into the local buffer for future Monte Carlo reward return.

Consequently, the \texttt{AttenAssemble} agent consistently operates over a candidate layer zoo from both its local base model and neighboring peers. Thus, the candidate layer zoo is formally defined as $\mathcal{Z}_k^{(i)}=\{(\theta_{k,l}^{(j)},m_{k}^{(j)})\mid j \in \{i\} \cup \mathcal{N}_i, l \in \{1, \dots, L_j\} \}$. While the size of the candidate pool remains stable, the underlying parameters of these modules are continuously refined, ensuring progressive knowledge assimilation without action space expansion.

\paragraph{Parameter Aggregation} While decentralized module sharing facilitates knowledge transfer, it cannot guarantee that the distributed knowledge will converge to a stable consensus throughout the collaboration. To ensure a progressive convergence, we introduce a source-based aggregation mechanism, as illustrated in Fig.~\ref{fig:FedJigsaw}. Following the local training, the parameters of the assembled layers are transmitted back to their originating source client for localized aggregation.

We use $\mathcal{P}_{k,l}^{(j)}$ to denote the subset of clients that borrowed module $l$ from source client $j$ during round $k$. The source client aggregates the updated parameters $\theta_{k,l}^{j \to p}$ from all borrowing peers $p \in \mathcal{P}_{k,l}^{(j)}$ following the Eq.~\ref{eq:aggregation}. In this way, the local updates will be consolidated into a unified consensus.

\begin{equation}
    \theta_{k+1, l}^{(j)} = \frac{1}{|\mathcal{P}_{k,l}^{(j)}|} \sum_{p \in \mathcal{P}_{k,l}^{(j)}} \theta_{k,l}^{(j \to p)}
    \label{eq:aggregation}
\end{equation}

\subsubsection{Non-Cooperative Game Challenge}
\label{sec:pomdp}
\paragraph{Partially Observable Challenge} When the model assembly is strictly confined to an isolated client, the sequential decision process is a standard MDP. However, the \texttt{SymbioArchitect} mechanism escalates system complexity, thereby casting the problem as a Partially Observable Markov Decision Process (POMDP). Due to strict privacy constraints, an agent cannot directly observe its neighbors' underlying data feature $E_{data}^{(j)}$ or model parameters, but is restricted to observing only the broadcasted structural metadata. Consequently, the local \texttt{AttenAssemble} agent faces severe partial observability: it must make critical model-assembling decisions without knowing the underlying data distributions that shaped the neighboring candidate modules.

Crucially, this partial observability is exacerbated by the multi-agent nature. As all clients concurrently update their local policies, the environment from the perspective of any individual agent becomes highly non-stationary. Consequently, the entire system evolves into a \textit{Non-Cooperative Game}. If each client acts greedily to maximize only its local validation accuracy, the system is prone to converging to a suboptimal Nash Equilibrium. Even if each individual exhibits exploration behavior, it still lacks the capability to guide the entire system from one Nash Equilibrium to another.

\paragraph{Implicit Collaboration Exploration via Reward Design:} Since a standard Nash Equilibrium is rarely Pareto-Optimal, blind and selfish local exploration is insufficient. To guide the entire multi-agent system towards a globally cooperative and stable state, clients must engage in \textit{implicit collaboration} during their exploration. To achieve this, we design the environment feedback by formulating the terminal reward $R_{k,T}^{(i)}$ as a weighted combination of local performance, a targeted exploration bonus, and hardware constraints:
\begin{equation}
    R_{k,T}^{(i)} = \alpha \cdot \text{Acc}^{(i)} + (1 - \alpha) \cdot \text{Exp}^{(i)} + \beta \cdot \text{Cap}^{(i)}
    \label{eq:reward}
\end{equation}

where $\text{Acc}^{(i)}$ represents the validation accuracy of the fully assembled model, and $\text{Cap}^{(i)}$ is the hardware utilization bonus, which guides the agent away from generating too shallow models. To break the suboptimal Nash trap, $\text{Exp}^{(i)}$ explicitly incentivizes clients with similar data distributions to discover diverse, orthogonal topological structures, defined as:
     
\begin{equation}
    \text{Exp}^{(i)} = \frac{1}{|\mathcal{N}| - 1} \sum_{j \neq i} \text{cos}(E_{data}^{(i)}, E_{data}^{(j)}) \cdot \left( 1 - \text{cos}(E_{arch}^{(i)}, E_{arch}^{(j)}) \right)
    \label{eq:exp_bonus}
\end{equation}

where $E_{arch}^{(i)}$ represents the encoded architectural feature of the assembled model. Intuitively, the higher the data similarity (cosine similarity of $E_{data}$) between two clients, the stronger the penalty if they adopt similar architectures. By forcing homogeneous clients to explore distinct branches of action space, this mechanism encourages a thorough exploration. Notably, the $\text{Exp}^{(i)}$ computation leverages the global coordination introduced by \texttt{Core-Tune} (Sec.~\ref{sec:coretune}). By evaluating data similarity solely on a task-agnostic auxiliary dataset, we ensure that the genuine feature representations $E_{data}$ of the local task data are never shared, in order to address privacy vulnerabilities.


\subsection{CoRe-Tune: Attention-Enhanced CTDE Tuning Strategy}
\label{sec:coretune}
As discussed in Sec.~\ref{sec:pomdp}, relying solely on decentralized policy updates within a POMDP inevitably leads to severe environmental non-stationarity. To break this information asymmetry, we propose \texttt{CoRe-Tune}, an attention-enhanced CTDE tuning strategy that utilizes a task-agnostic proxy dataset on the clients to explicitly guide the local agents' policies. By doing so, we successfully impart a global vision to the agents while rigorously preserving privacy, as the genuine local data distributions are never exposed to any party within the MHFL system. Recognizing that access to a suitable proxy dataset is rarely guaranteed in practical federated deployments, we deliberately design this server-side guidance as an \textbf{Optional Phase}. The detailed pseudocode is provided in Appendix \ref{app:fedjigsaw}.


\texttt{CoRe-Tune} uses a CTDE scheme to introduce a centralized critic network parameterized by $\vartheta$ on the server. This scheme will ask local clients to share information during policy training, but once the policies converge, the server can be \textbf{fully decoupled}. This ensures that during the execution phase, clients can make assembly decisions independently, based solely on their local observations, while still maintaining the cooperative behaviors learned from the centralized critic.

As shown in Fig.~\ref{fig:FedJigsaw}, the centralized server collects the local data feature $E_{data}^{(i)}$, which encapsulates the local data distribution and the assembled model characteristics through the transformer encoding in \texttt{AttenAssemble}. These features are processed by a Transformer Encoder to formulate the global information $Z_t$. Meanwhile, the local observation history $h_t^{(i)}$ is fed into another Transformer Encoder to generate the history embedding $E_{History,t}^{(i)}$. Although the actor's decision relies on current observations, the critic must be conditioned on the observation history. As highlighted by Lyu \textit{et al.}~\cite{lyu2023centralized}, without an observation history sequence, the critic would have less information than the actor, leading to biased value estimations and failing to infer the policy accurately. The local policy network $\theta_i$ is then optimized by minimizing the following global policy gradient objective:
\begin{equation}
    \nabla_{\theta_i} J(\theta_i) = \mathbb{E}_{\pi} \left[ \sum_{t=1}^{T_i} \nabla_{\theta_i} \log \pi_{\theta_i}(a_{k,t}^{(i)} \mid h_{k,t}^{(i)}) A_{\vartheta}(h_{k,t}^{(i)}, Z_t) \right]
    \label{eq:policyupdate}
\end{equation}

where $h_{k,t}^{(i)}$ represents the local observation history of client $i$ up to step $t$, aligning with the decentralized execution constraint. Crucially, the advantage function $A_{\vartheta}(h_{k,t}^{(i)}, Z_t) = r_{k,t}^{(i)} - V_{\vartheta}(h_{k,t}^{(i)}, Z_t)$ is calculated by subtracting the centralized critic's value estimation from the empirical Monte Carlo return, which we have discussed in Sec.~\ref{sec:MDP}. The detailed experiment setup and convergence analysis is detailed in App.~\ref{app:exp_advanced}.

\section{Experiments}
\subsection{Experiment Setup}

\paragraph{Experimental Setup \& Baselines:} 
We evaluate \texttt{FedJigsaw} on five image classification datasets (CIFAR-10, CIFAR-100, Tiny-ImageNet, FEMNIST, and SVHN), partitioning the data across 10 continuously active clients organized in a fully connected topology using a Dirichlet distribution $\textit{Dir}(\sigma)$ to simulate Non-IID scenarios \citep{FedShapleX,pfedhr}. To model system heterogeneity, clients are assigned diverse base architectures: lightweight CNNs (2, 3, 4, or 6 convolutional layers followed by 2 FC layers) for simpler datasets, and VGG variants (VGG-11, 13, 16, 19) for complex tasks. We benchmark against state-of-the-art Partial Training (PT) and model reassembly methods—\textbf{HeteroFL~\citep{diao2020heterofl}}, \textbf{FedRolex~\citep{alam2022fedrolex}}, \textbf{FedShapleX~\citep{FedShapleX}}, and \textbf{pFedHR~\citep{pfedhr}}. For a fair comparison, the largest target architecture in each setting is consistently initialized as the global supernet across all baselines.


\subsection{Performance Validation}
\paragraph{Model Performance:}Tab.~\ref{tab:main_accuracy} demonstrates that FedJigsaw achieves superior personalization capacity across MHFL systems, yielding an average absolute \textbf{local accuracy improvement} of 3.94\% (13.87\% relative) across all settings. It delivers state-of-the-art results on complex tasks under High Non-IID scenarios, reaching 84.93\% on CIFAR-10, 53.77\% on CIFAR-100, and 35.14\% on Tiny-ImageNet, significantly outperforming the best baselines (62.30\%, 50.56\%, and 21.79\%, respectively). In Low Non-IID scenarios where local datasets are class-rich and uniformly distributed, PT-based methods occasionally show slight advantages. This is theoretically expected, as their shared global backbone acts as a strong prior to mitigate the local capacity bottlenecks of small models. However, despite being a decentralized backbone-free approach lacking such global guidance, FedJigsaw successfully maintains highly competitive and stable performance, effectively avoiding the catastrophic memory failures (e.g., OOM) prevalent in centralized NAS methods.

Beyond raw accuracy, the \textbf{variance} metrics reveal FedJigsaw's exceptional stability and cross-client fairness. PT-based methods often exhibit performance fluctuations across clients (e.g., FedShapleX reaching a massive 0.1877 variance on FEMNIST, while HeteroFL hitting 0.1269). This instability occurs because forcing a global backbone onto highly heterogeneous local data through subnet extractions inevitably benefits some clients while penalizing others. In contrast, FedJigsaw maintains a consistently tight variance across almost all scenarios (e.g., restricting the variance to 0.0431 on FEMNIST on High Non-IID). This proves that our decentralized dynamic assembly allows every individual client to tailor an optimal architecture for its specific data distribution, ensuring that high performance is shared fairly rather than skewed towards a few specific clients.

\begin{table*}[t]
\centering
\resizebox{\textwidth}{!}{
\begin{tabular}{l|cc|cc|cc|cc|cc}
\toprule
\multirow{2}{*}{\textbf{Method}} & \multicolumn{2}{c|}{\textbf{CIFAR-10}} & \multicolumn{2}{c|}{\textbf{CIFAR-100}} & \multicolumn{2}{c|}{\textbf{Tiny-ImageNet}} & \multicolumn{2}{c|}{\textbf{FEMNIST}} & \multicolumn{2}{c}{\textbf{SVHN}} \\
\cmidrule(lr){2-3} \cmidrule(lr){4-5} \cmidrule(lr){6-7} \cmidrule(lr){8-9} \cmidrule(lr){10-11}
& Low & High & Low & High & Low & High & Low & High & Low & High \\
\midrule
\multicolumn{11}{l}{\textit{PT-based MHFL}} \\
\midrule
HeteroFL & 49.40\tiny{$\pm$0.0567} & 54.05\tiny{$\pm$0.0464} & 37.42\tiny{$\pm$0.0217} & 31.19\tiny{$\pm$0.0275} & 14.39\tiny{$\pm$0.0179} & 20.10\tiny{$\pm$0.0393} & 61.07\tiny{$\pm$0.1269} & 75.28\tiny{$\pm$0.0523} & \textbf{86.11}\tiny{$\pm$0.0326} & 84.95\tiny{$\pm$0.0880} \\
FedRolex & 47.80\tiny{$\pm$0.0556} & 61.60\tiny{$\pm$0.0529} & 36.75\tiny{$\pm$0.0420} & 37.34\tiny{$\pm$0.0324} & 15.30\tiny{$\pm$0.0198} & 20.89\tiny{$\pm$0.0310} & 67.26\tiny{$\pm$0.1055} & 74.47\tiny{$\pm$0.0624} & 84.14\tiny{$\pm$0.0417} & 85.49\tiny{$\pm$0.0678} \\
FedShapleX & 54.03\tiny{$\pm$0.0713} & 62.30\tiny{$\pm$0.0519} & 27.09\tiny{$\pm$0.0217} & 50.56\tiny{$\pm$0.0528} & 14.54\tiny{$\pm$0.0162} & 21.79\tiny{$\pm$0.0351} & \textbf{80.11}\tiny{$\pm$0.0618} & 74.58\tiny{$\pm$0.1877} & 83.35\tiny{$\pm$0.0486} & 88.31\tiny{$\pm$0.0819} \\
\midrule
\multicolumn{11}{l}{\textit{NAS-based FL}} \\
\midrule
pFedHR & 45.45\tiny{$\pm$0.0054} & 54.16\tiny{$\pm$0.1379} & OOM & OOM & OOM & OOM & 70.78\tiny{$\pm$0.0309} & 73.77\tiny{$\pm$0.0962} & 78.23\tiny{$\pm$0.0197} & 83.55\tiny{$\pm$0.0769} \\
\textbf{FedJigsaw (Ours)} & \textbf{59.61}\tiny{$\pm$0.0543} & \textbf{84.93}\tiny{$\pm$0.0616} & \textbf{37.43}\tiny{$\pm$0.0208} & \textbf{53.77}\tiny{$\pm$0.0394} & \textbf{21.10}\tiny{$\pm$0.0391} & \textbf{35.14}\tiny{$\pm$0.0340} & 73.01\tiny{$\pm$0.0464} & \textbf{76.72}\tiny{$\pm$0.0431} & 78.26\tiny{$\pm$0.0449} & \textbf{90.59}\tiny{$\pm$0.0567} \\
\midrule
\multicolumn{11}{l}{\textit{Ablation Study}} \\
\midrule
W/O SymbioArchitect & 51.83\tiny{$\pm$0.0797} & 81.74\tiny{$\pm$0.0901} & 27.55\tiny{$\pm$0.0411} & 47.87\tiny{$\pm$0.0540} & 13.17\tiny{$\pm$0.0309} & 29.07\tiny{$\pm$0.0347} & 69.77\tiny{$\pm$0.0519} & 76.46\tiny{$\pm$0.0563} & 74.97\tiny{$\pm$0.0779} & 86.32\tiny{$\pm$0.1069} \\
W/O CoRe-Tune      & 52.57\tiny{$\pm$0.0711} & 82.24\tiny{$\pm$0.0803} & 33.51\tiny{$\pm$0.0268} & 50.97\tiny{$\pm$0.0477} & 18.92\tiny{$\pm$0.0317} & 31.21\tiny{$\pm$0.0513} &69.26\tiny{$\pm$0.0451} & 76.16\tiny{$\pm$0.0532} & 72.36\tiny{$\pm$0.1088} & 86.03\tiny{$\pm$0.1216} \\
\bottomrule
\end{tabular}
}
\caption{Test accuracy (\%) and standard deviation (across clients) across five datasets under Low Non-IID (e.g., $\alpha=1.0$) and High Non-IID (e.g., $\alpha=0.1$) settings. OOM stands for Out Of Memory.}
\label{tab:main_accuracy}
\end{table*}

\paragraph{Decision-Making Efficiency:} Tab.~\ref{tab:efficiency_overhead} compares the system overhead regarding peak memory usage (RAM) and latency (Time) of the decision-making component, excluding local training. FedJigsaw demonstrates exceptional computational efficiency among policy-based methods, drastically reducing the latency to 1.11–1.53 seconds. This is an order of magnitude faster than FedShapleX and highly competitive with extremely lightweight rule-based heuristic baselines. Regarding memory footprint, while FedJigsaw requires moderately higher RAM on the centralized server when using CoRe-Tune compared to heuristic approaches, it introduces minimal additional memory and computational costs on the local clients. However, FedJigsaw's memory requirement remains well within practical hardware limits and completely avoids Out-Of-Memory (OOM) failures that paralyze NAS or model reassembly methods like pFedHR, proving its outstanding scalability.

\begin{table*}[t]
\centering
\resizebox{0.8\textwidth}{!}{
\begin{tabular}{l|cc|cc|cc|cc|cc}
\toprule
\multirow{2}{*}{\textbf{Method}} & \multicolumn{2}{c|}{\textbf{CIFAR-10}} & \multicolumn{2}{c|}{\textbf{CIFAR-100}} & \multicolumn{2}{c|}{\textbf{Tiny-ImageNet}} & \multicolumn{2}{c|}{\textbf{FEMNIST}} & \multicolumn{2}{c}{\textbf{SVHN}} \\
\cmidrule(lr){2-3} \cmidrule(lr){4-5} \cmidrule(lr){6-7} \cmidrule(lr){8-9} \cmidrule(lr){10-11}
& RAM & Time & RAM & Time & RAM & Time & RAM & Time & RAM & Time \\
\midrule
\multicolumn{11}{l}{\textit{Heuristic-based MHFL}} \\
\midrule
HeteroFL & 0.58 & 0.72 & 0.59 & 1.13 & 0.52 & 1.71 & 0.26 & 1.34 & 0.27 & 1.97 \\
FedRolex & 0.57 & 0.98 & 0.63 & 1.39 & 0.57 & 1.27 & 0.29 & 1.76 & 0.29 & 1.03 \\
\midrule
\multicolumn{11}{l}{\textit{Policy-based MHFL}} \\
\midrule
FedShapleX & \textbf{1.81} & 9.61 & \textbf{1.93} & 11.90 & \textbf{3.76} & 19.34 & 2.69 & 11.04 & \textbf{2.36} & 13.88 \\
pFedHR & 61.70 & 626.71 & OOM & OOM & OOM & OOM & 37.20 & 1057.32 & 23.09 & 1152.60 \\
\textbf{FedJigsaw (Ours)} & 2.61 & \textbf{1.11} & 7.65 & \textbf{1.29} & 7.69 & \textbf{1.53} & \textbf{1.26} & \textbf{1.28} & 2.59 & \textbf{1.40} \\
\bottomrule
\end{tabular}
}
\caption{System overhead comparison. RAM denotes the peak local memory footprint (in GB), and Time represents the average computation/communication latency per round (in seconds). }
\vspace{-10pt}

\label{tab:efficiency_overhead}
\end{table*}

\subsection{Ablation Studies}

\paragraph{Effectiveness of \texttt{SymbioArchitect}:} 
To confirm that models genuinely share knowledge rather than merely overfitting locally, we evaluate the variant enabling only \texttt{SymbioArchitect} (W/O CoRe-Tune) against the strictly isolated baseline (W/O SymbioArchitect). As shown in Tab.~\ref{tab:main_accuracy}, \texttt{SymbioArchitect} delivers consistent improvements across all benchmarks. Under Low Non-IID, it notably boosts local accuracy on CIFAR-100 (27.55\% $\rightarrow$ 33.51\%) and Tiny-ImageNet (13.17\% $\rightarrow$ 18.92\%). These gains prove that decentralized module sharing effectively shatters local data isolation, empowering clients to acquire missing class representations from their topological peers in severely Non-IID environments.


\paragraph{Effectiveness of \texttt{CoRe-Tune}:} 
Comparing the full framework against the W/O CoRe-Tune variant demonstrates that \texttt{CoRe-Tune} significantly enhances accuracy and cross-client stability. Mitigating the severe environmental non-stationarity inherent in decentralized MARL prevents catastrophic strategy oscillation. For instance, \texttt{CoRe-Tune} boosts CIFAR-100 accuracy from 33.51\%/50.97\% to 37.43\%/53.77\% (Low/High Non-IID), while shrinking variance on SVHN from 0.1088/0.1216 to 0.0449/0.0567. Notably, even if \texttt{CoRe-Tune} is disabled due to no available proxy dataset, \texttt{FedJigsaw} remains highly competitive. On the challenging High Non-IID Tiny-ImageNet, the W/O CoRe-Tune variant still achieves 31.21\% accuracy, surpassing the best baseline (FedShapleX at 21.79\%) by nearly 10\%. Overall, \texttt{FedJigsaw} delivers strong standalone performance without auxiliary data, while \texttt{CoRe-Tune} further pushes its boundaries to achieve state-of-the-art collaborative efficiency.

\section{Conclusion}
In this paper, we proposed \texttt{FedJigsaw}, a novel framework that reformulates PT-based MHFL from rigid sub-network extraction into a multi-agent decentralized model reassembly problem. We introduced \texttt{AttenAssemble}, an attention-based sequential decision paradigm empowering clients to autonomously construct customized architectures with minimal local latency. To break the physical isolation of heterogeneous backbones, we designed \texttt{SymbioArchitect}, elevating knowledge transfer from parameter aggregation to structural module sharing within topological neighborhoods. Furthermore, to mitigate the non-stationarity of multi-agent interactions, we developed \texttt{CoRe-Tune}, an attention-enhanced tuning strategy that stabilizes local policies while preserving privacy. Extensive experiments demonstrate that \texttt{FedJigsaw} significantly improves local accuracy and cross-client fairness while maintaining an exceptionally lightweight memory footprint, offering a scalable solution for resource-constrained edge deployments. The limitations are discussed in Appendix~\ref{app:limitations}.

\bibliographystyle{iclr2027_conference}
\bibliography{references}

\newpage
\appendix

\section{Technical appendices and supplementary material}

\subsection{AttenAssemble Model Assembly Pseudocode}
\label{app:actor}

Algorithm~\ref{alg:atten_assemble} describes the action flow of the proposed \texttt{AttenAssemble} module, which dynamically constructs a customized neural block by block. The assembly loop (steps 1-5) iteratively builds the model sequence. It fundamentally consists of State Encoding (lines 2-7), which encodes the current state into a sequence, Valid Action Collecting (lines 8-11), which collects the valid metadata, Action Encoding (lines 13-15), which encodes the valid actions into a sequence, and Transformer-based policy execution (lines 18-20). Afterward, the generated abstract metadata sequence is mapped and instantiated as a physical neural network suitable for local training (lines 21-25).

\begin{algorithm}[h]
\caption{\texttt{AttenAssemble}: Attention-based Sequential Architecture Assembly}
\label{alg:atten_assemble}
\KwIn{Client physical capacity $C_i$, Local data embedding $\mathcal{E}_{data}$, 
Topological neighbors metadata pool $\mathcal{P}_i$, Actor network $\theta_i$, Layer encoder $\mathcal{E}_{layer}$, Temperature $\tau$.}

\KwOut{Assembled customized model $M_i$, RL trajectory $\mathcal{T}_i$.}

\textbf{Initialization: } State sequence $\mathcal{S} \leftarrow \emptyset$, RL Trajectory $\mathcal{T}_i \leftarrow \emptyset$, Transformer token sequence $\mathbf{E} \leftarrow [\mathcal{E}_{data}]$, Current output dimension $d_{out} \leftarrow \text{Initial Shape}$\;

\For{step $t = 1, 2, \dots, C_i$}{
    \tcp{1. State Encoding}
    \If{$\mathcal{S}$ is not empty}{
        \For{each metadata $m \in \mathcal{S}$}{
            Encode state token: $enc_{s} \leftarrow \mathcal{E}_{layer}(m, \text{segment=1})$\;
            Append $enc_{s}$ to token sequence $\mathbf{E}$\;
            Update $d_{out}$ according to $m$\;
        }
    }
    
    \tcp{2. Valid Action Collecting}
    Initialize valid action set $\mathcal{A}_{valid} \leftarrow \emptyset$\;
    \For{each metadata $m \in \mathcal{P}_i$}{
        \If{(VolumeMatch($m, d_{out}$))}{
            $\mathcal{A}_{valid} \leftarrow \mathcal{A}_{valid} \cup \{m\}$\;
        }
    }
    $\mathcal{A}_{valid} \leftarrow \mathcal{A}_{valid} \cup \{\text{STOP\_TOKEN}\}$\;
    
    \tcp{3. Action Encoding}
    \For{each $a \in \mathcal{A}_{valid}$}{
        Encode action token: $enc_{a} \leftarrow \mathcal{E}_{layer}(a, \text{segment=2})$\;
        Append $enc_{a}$ to token sequence $\mathbf{E}$\;
    }
    
    \tcp{4. Pointer Network Decision}
    $\mathbf{P} \leftarrow Transformer(\mathbf{E},\tau,\pi_{\theta})$\;
    Sample action $a_t \sim \text{Categorical}(\mathbf{P})$\;
    
    \tcp{5. State Update \& Trajectory Recording}
    Append $a_t$ to $\mathcal{S}$\;
    Record $(\log p(a_t), a_t, \mathbf{z}_{global})$ into trajectory $\mathcal{T}_i$\;
    
    \lIf{$a_t == \{\text{STOP\_TOKEN}\}$}{\textbf{break}}
}

\tcp{6. Model Assembly}
Initialize empty neural sequential container $\mathcal{M}_i$\;
\For{each module $m \in \mathcal{S} \setminus \{\text{STOP\_TOKEN}\}$}{
    Fetch physical layer $\theta_m$ \;
    Append $\theta_m$ to $\mathcal{M}_i$\;
}

\Return $\mathcal{M}_i, \mathcal{T}_i$\;
\end{algorithm}

\subsection{FedJigsaw MHFL Framework Pseudocode}
\label{app:fedjigsaw}
Algorithm~\ref{alg:fedjigsaw} describes the detailed procedure of the proposed \texttt{FedJigsaw} framework, which includes dynamic model assembly, local training, and multi-agent policy optimization. The client first assembles the local model using algorithm~\ref{alg:atten_assemble} (lines 5-6). Then, during local training, the assembled model $\mathcal{M}_{i,k}$ will be updated on the local dataset and evaluated (lines 7-9). Afterward, using the Actor-Critic Policy, the final reward is first computed, and then the advantage function is computed from the reward and the critic's estimated $V_{\vartheta}$ (lines 10-12). As the reward is computed, the actor $\theta_i$ is updated and the critic loss is recorded (lines 13-14). Then, as the models are assembled and trained, the updated parameters will be sent back for aggregation (lines 15-17). Finally, the critic $V_\vartheta$ will be updated using the recorded critic loss (line 18).


\begin{algorithm}[h]
\caption{\texttt{FedJigsaw}}
\label{alg:fedjigsaw}
\KwIn{Communication rounds $K$, Client set $\mathcal{N}$, Client datasets $\{\mathcal{D}_i\}_{i=1}^{|\mathcal{N}|}$, Topology graph $\mathcal{G}$, Server Critic $V_\phi$, Global module pool $\Omega$.}
\KwOut{Optimized local architectures and corresponding trained modules.}

\textbf{Server Initialization: } Initialize global modular weights in $\Omega$, shared encoders $\mathcal{E}_{layer}, \mathcal{E}_{data}$, and Server Critic $V_\phi$\;
\textbf{Client Initialization: } Each client $i \in \mathcal{N}$ initializes Actor $\pi_{\theta_i}$ and topology neighborhood $\mathcal{P}_i$ based on $\mathcal{G}$\;

\For{round $k = 1, 2, \dots, K$}{
    \For{each client $i \in S_k$}{
        \tcp{1. Local Architecture Assembly (Refer to Alg. 1)}
        Execute \texttt{AttenAssemble} to obtain model and trajectory: \\
        $(\mathcal{M}_{i,k}, \mathcal{T}_{i,k}) \leftarrow \text{Algorithm 1}(C_i, \mathcal{E}_{data}(\mathcal{D}_i), \mathcal{P}_i, \pi_{\theta_i}, \mathcal{E}_{layer}, \tau)$\;
        
        \tcp{2. Local Training \& Evaluation}
        Train assembled model $\mathcal{M}_{i,k}$ on local dataset $\mathcal{D}_i$ for $E$ epochs\;
        Evaluate model to obtain test accuracy $Acc_{i,k}$\;
        Extract updated parameters $\Delta w_{i,k}$ of the selected modules\;
        
        \tcp{3. Actor-Critic Policy Optimization}
        Compute final reward $R_{i,k} \leftarrow \alpha Acc_{i,k} + (1-\alpha)Exp_i + \beta Cap_i$\;
        \For{each step $t$ in trajectory $\mathcal{T}_{i,k}$}{
            Compute Advantage $A_t \leftarrow \gamma^{T-t} R_{i,k} - V_\vartheta(s_t, \mathbf{z}_{global})$\;
            Update Actor $\theta_i$ via $\nabla_{\theta_i} \log \pi_{\theta_i}(a_t|s_t) A_t$\;
            Calculate Critic Loss $\mathcal{L}_{\phi} \leftarrow \text{MSE}(v_s, G_t)$\;
        }
    }
    \For{each structural module $m \in \Omega$}{
        \If{module $m$ is selected by any client in $S_k$}{
            Parameter Aggregation: $w_m \leftarrow w_m + \frac{1}{|S_{k,m}|} \sum_{i \in S_{k,m}} \Delta w_{i, m}$\;
        }
    }
    Update Server Critic $V_\vartheta$ using aggregated $\mathcal{L}_{\phi}$\;
}
\end{algorithm}

\subsection{Further discussion on Submodel Slicing in PT-based MHFL}

While Partial Training paradigms, such as HeteroFL \citep{diao2020heterofl} and FedRolex \citep{alam2022fedrolex}, have pioneered the mitigation of computational heterogeneity. Their reliance on rigid \textit{submodel slicing} introduces several fundamental architectural limitations.

\paragraph{Semantic Misalignment and Residual Coupling Breakdown}
In typical PT-based methods, sub-networks are generated by aggressively selecting neurons from each hidden layer and truncating the width of the global model. However, modern neural architectures rely heavily on complex structural couplings, such as residual connections.

When a global supernet is arbitrarily sliced into narrower heterogeneous sub-networks, the strict dimensional equivalence required for addition operations is broken. As addition operations across different sub-models may be affected by semantic misalignment, this can lead to severe \textit{semantic drift} and undermine the mathematical integrity of the original feature maps. By contrast, \texttt{FedJigsaw} completely circumvents this by operating at the granularity of cohesive functional modules. This ensures that internal mathematical constraints are perfectly preserved during the reassembly process.

\paragraph{Topological Homogenization}
A more serious limitation of submodel slicing is its inherent inability to foster true architectural diversity. Because every client's model is merely a scaled-down projection of the exact same global supernet, the topological search space is strictly bounded by the supernet's structural template. A sliced VGG network remains topologically the same as the original VGG, but with a different width. Consequently, PT methods suffer from topological homogenization. They assume that a smaller version of a universally well-performing architecture is inherently optimal for resource-constrained clients. However, under severe Non-IID conditions, a client might actually benefit from a different topology. 

\texttt{FedJigsaw} tries to mitigate the "Supernet Trap" issue by replacing the top-down slicing with a bottom-up model assembly. Instead of destroying a large network, clients construct personalized local topologies sequentially using the \texttt{AttenAssemble} paradigm. Although the assembly is still limited to sequential embedding, \texttt{FedJigsaw} grants clients the freedom to explore an exponentially larger, structurally diverse combinatorial space.

\subsection{Implementation details and Further analysis}
\label{app:exp_advanced}

We implement \texttt{FedJigsaw} using PyTorch 2.10.0 and simulate the MHFL environment on a dedicated server equipped with an NVIDIA RTX 6000 Ada GPU (256GB VRAM) and a 128-core CPU. All local models are optimized using the Adam optimizer with a learning rate of 0.1 for 1000 policy training rounds (with \texttt{CoRe-Tune}) and another 1000 model training rounds. To balance exploration and exploitation, the hyperparameter $\alpha$ is set to 0.7 and 0.9 for the Low and High Non-IID settings, respectively.

\begin{figure}[ht]
    \centering
    \includegraphics[width=0.8\linewidth]{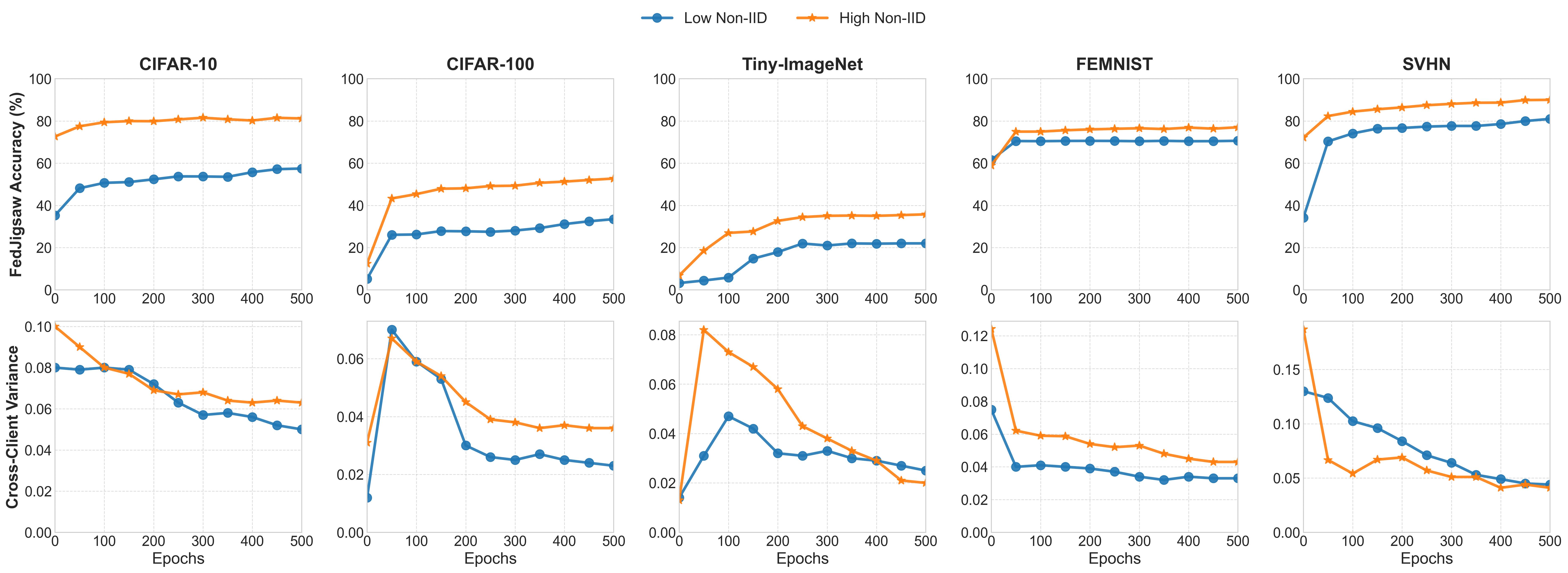}
    \caption{Learning dynamics of \texttt{FedJigsaw} across five datasets over 500 communication rounds. The top row presents the convergence of test accuracy, while the bottom row illustrates the cross-client performance variance under both Low Non-IID and High Non-IID settings.}
    \label{fig:model_convergence}
\end{figure}

\paragraph{Model Convergence:} Figure~\ref{fig:model_convergence} illustrates the learning dynamics of \texttt{FedJigsaw} across five diverse datasets over 500 epochs. As shown in the top row, \texttt{FedJigsaw} achieves stable and continuous convergence in test accuracy under both Low and High Non-IID settings. Notably, the framework exhibits exceptional resilience to severe data heterogeneity, maintaining robust and even superior performance in High Non-IID scenarios. More importantly, the bottom row validates the systemic stability of our multi-agent framework. While the cross-client performance variance exhibits initial fluctuations during the early architectural exploration phase, it consistently and sharply declines, stabilizing at a remarkably low level across all datasets. This continuous variance reduction confirms that our decentralized module sharing and \texttt{CoRe-Tune} mechanisms successfully foster implicit collaboration. Instead of diverging due to local data biases, the participating agents effectively mitigate individual capacity bottlenecks and converge toward a stable, globally cooperative state with high cross-client fairness.

\begin{figure}[ht]
    \centering
    \includegraphics[width=0.8\linewidth]{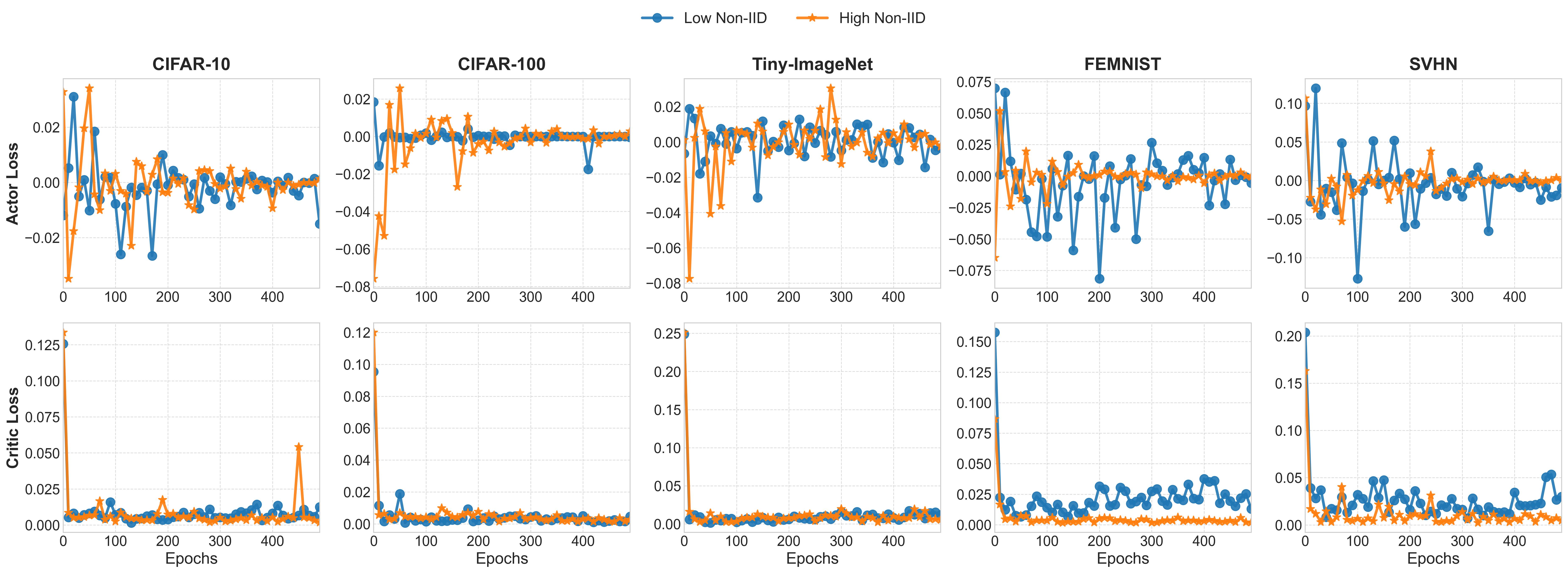}
    \caption{Training dynamics of the Actor (top) and Critic (bottom) losses over 500 epochs.}
    \label{fig:a2c_loss}
\end{figure}

\paragraph{Policy Convergence:}
Figure~\ref{fig:a2c_loss} visualizes the training trajectories of the \texttt{CoRe-Tune} networks over 500 epochs. Across all datasets, the Critic loss rapidly converges near zero, while the Actor loss naturally oscillates around zero with diminishing amplitude. Notably, \texttt{FedJigsaw} optimizes more efficiently under High Non-IID conditions than Low Non-IID. For instance, in FEMNIST and SVHN, the Low Non-IID setup (blue lines) exhibits larger Actor oscillations and higher residual Critic errors. This is because homogeneous data (Low Non-IID) constrains state exploration, whereas extreme heterogeneity (High Non-IID) acts as an implicit regularizer. This diverse structural variation provides the central Critic with a richer state space, accelerating global value consensus and leading to tighter, more stable policy convergence.

\subsection{Limitations}
\label{app:limitations}

While \texttt{FedJigsaw} introduces a promising decentralized module-assembly paradigm for MHFL, we acknowledge several limitations that pave the way for future research.

\paragraph{Manual Module Specification:} Currently, to preserve semantic integrity and simplify the state space, \texttt{FedJigsaw} relies on human-designed functional blocks (e.g., ResNet or Transformer Blocks) as the basic unit of assembly. This manual uses automated graph-partitioning algorithms to dynamically discover and slice fine-grained modules, thereby eliminating human intervention.

\paragraph{Sequential Topology and Residual Connection Constraints:} The current \texttt{AttenAssemble} paradigm primarily operates under a sequential, linear assembly constraint. While this efficiently circumvents dimensional mismatches, it precludes the generation of highly complex architectures with intricate cross-module skip connections (e.g., DenseNet-like topologies). Developing a topology-aware alignment mechanism is a critical next step. In the future, the bottom-up model assembly should evolve to a topology generation problem.

\paragraph{Reactive Exchange and Proactive Knowledge Diffusion:} Although \texttt{SymbioArchitect} successfully facilitates structural knowledge transfer within topological neighborhoods, this exchange is fundamentally reactive, driven solely by local clients' autonomous assembly demands. This framework currently lacks a systemic mechanism to actively orchestrate knowledge propagation across the MHFL system. Future work will explore proactive diffusion strategies to consciously guide and accelerate the spread of critical knowledge across the entirely decentralized network. Hopefully, as this mechanism is guiding the knowledge diffusion across the client, we expect that a generalized model concluding all local knowledge could also be generated besides the local personalized models.

\subsection{Broader Impact}
\label{app:broaderimpacts}

The broader impact of \texttt{FedJigsaw} primarily lies in lowering the cost of deploying Federated Learning. Traditional centralized approaches usually depend on a powerful centralized server for coordination, which is unfriendly to start-up companies and individual developers. By reforming model personalization into a decentralized modular assembly problem, our framework eliminates the reliance on a computationally prohibitive server for supernet maintenance or architecture personalization for the clients. This lowers the barriers for deploying distributed AI, enabling resource-constrained edge devices, such as smartphones or IoT sensors, to actively participate in collaborative learning and promoting a more inclusive AI ecosystem. However, without strict centralized oversight, the exchange of modules could make the system vulnerable to Byzantine attacks, where malicious clients might broadcast poisoned modules across neighborhoods. Future deployment must incorporate robust, module-level anomaly detection to mitigate cyberattack risks.



\end{document}